\documentclass[runningheads]{llncs}
\usepackage{graphicx}
\usepackage{float}
\usepackage{xcolor}

\usepackage{url}
\makeatletter
\g@addto@macro{\UrlBreaks}{\UrlOrds}
\makeatother

\begin{document}
\title{Data Mining and Visualization to Understand Accident-prone Areas}
%
%

\author{Md. Mashfiq Rizvee\inst{1}\orcidID{0000-0001-7414-7951} \and
Md Amiruzzaman\thanks{corresponding author 1: \email{mamiruzz@kent.edu}}\inst{2}\orcidID{0000-0002-2292-5798} \and
Md. Rajibul Islam\thanks{corresponding author 2: \email{md.rajibul.islam@uap-bd.edu}}\inst{3}\orcidID{0000-0003-0565-6917}}
\authorrunning{Rizvee et al.}
%
\institute{North South University, Dhaka 1229, Bangladesh\\
\email{mashfiq.rizvee@northsouth.edu}
\and
Kent State University, OH 44242, USA\\
\email{mamiruzz@kent.edu}\\
\and
University of Asia Pacific, Dhaka 1205, Bangladesh\\
\email{md.rajibul.islam@uap-bd.edu}}
%
\maketitle              
%
\begin{abstract}
In this study, we present both data mining and information visualization techniques to identify accident-prone areas, most accident-prone time, day, and month. Also, we surveyed among volunteers to understand which visualization techniques help non-expert users to understand the findings better. Findings of this study suggest that most accidents occur in the dusk (i.e., between 6 to 7 pm), and on Fridays. Results also suggest that most accidents occurred in October, which is a popular month for tourism. These findings are consistent with social information and can help policymakers, residents, tourists, and other law enforcement agencies. This study can be extended to draw broader implications.

\keywords{cluster analysis \and correlation \and data mining \and information visualization \and user study.}
\end{abstract}
\section{Introduction}

Injuries and loss of life caused by road traffic crashes are a global problem and remarkably affect socioeconomic growth and social prosperity \cite{gaarder2004impact}. Road traffic injuries are estimated to be the eighth driving reason for death worldwide for all age groups and the main motive of death for kids and youngsters 5–29 years old. Road traffic crashes are causing a projected 1.35 million passing, and daily, right around 3,700 individuals are put to death globally in road traffic accidents associated with buses, cars, bicycles, motorcycles, vans, or people on foot \cite{world2018global}. In the United States, road traffic crashes are a leading purpose of dying for humans aged 1–54 and the preeminent source of non-normal demise for sound U.S. residents dwelling or touring abroad \cite{sauber_schatz_parker_sleet_ballesteros_2019}. Various vehicles and transportation such as trucks, bikes, mopeds, people on foot, creatures, taxis, and different voyagers are the users of roads all through the world. Travel made conceivable by motor vehicles helps monetary and social advancement in numerous nations. Yet every year, vehicles associated with crashes are accountable for many passing and injuries. 

Road and traffic crashes are characterized by a lot of factors that are for most of the discrete nature \cite{amiruzzaman2018prediction}. The serious issue in the investigation of crashes information is its heterogeneous nature \cite{savolainen2011statistical}. Although, investigators utilized segmentation of the crashes data to diminish this heterogeneity applying several estimates, for example, expert knowledge, yet there is no assurance that this will produce an optimum segmentation which comprises homogeneous sets of road crashes \cite{depaire2008traffic}. For that reason, cluster investigation which is a significant data mining technique can help the segmentation of road and traffic crashes. Such can be utilized as the earliest assignment to accomplish numerous objectives.

Road and traffic crashes are sometimes found to be increasingly repeated in certain specific areas. The investigation of these areas can assist in understanding certain features of accident occurrences that make a road crash to take place repeatedly in these areas. Interestingly, time analysis is found to be another significant road accident feature to understand accident-prone areas. Various data mining techniques have been used in literature to conduct these analyses. Data mining utilizes various algorithms and procedures to find the correlation in a huge volume of data which is viewed as one of the most significant tools. DBSCAN algorithm is one of the common density-based clustering algorithms of data mining that makes a group of abstract objects into classes of similar objects. This clustering algorithm is used in our study on pedestrian crashes data that has been taken from the U.S. Government’s national database for public use (i.e., https://www.data.gov  \cite{usdatagov}). The quality of the clusters obtained by DBSCAN is then verified using a heatmap analysis. A past study indicated the importance of time and day which can be helpful to understand which days and times have a higher chance of accident \cite{amiruzzaman2018prediction}. So, in this study, we ran statistical analysis and from the descriptive statistics, we found which month, day, and times are highly accident-prone. 

The purpose of this study was three folds. First, to understand and identify the accident-prone places based on pedestrian crashes data. Second, identify most accident-prone time, day, and month which could help pedestrians to be more cautious during those times. Third, find appropriate visual techniques to present the findings so that both non-experts (e.g., pedestrian) and experts could comprehend the information better.


Below we list contributions of this study:

\begin{itemize}
    \item We identify the accident-prone areas using the Density-Based Spatial Clustering of Application with Noise (DBSCAN) clustering algorithm.  
    \item We generate a ``Heatmap'' on a map showing the number of accidents that happened between January to December. The heatmap shows which places had the most number of accidents. 
    \item We further analyze the data by breaking it down to the most accident-prone month, day, hour, and location using different visualization techniques.
    \item We conducted a user study survey to understand which visualization figures help to understand the findings better. 
\end{itemize} 

\section{Related work}
Literature-based on road and traffic accidents has been of interest to many. For example, Kumar and Toshniwal \cite{kumar2015data} analyzed the main factors associated with a road and traffic accident. In their study, the authors used association mining rules to identify the circumstances that are correlated with the occurrence of each accident. As the number of road accidents is increasing day by day, road and traffic safety is becoming a priority. Therefore, a study could focus on finding ways to improve road traffic safety and manage it efficiently. 

Tian, Yang, and Zhang \cite{tian2010method} analyzed the causes of road accidents based on data mining, and provided a method of traffic data analysis which can improve road traffic safety management effectively. Continuing the journey to find the factors that make more damages in road accidents. Amiruzzaman \cite{amiruzzaman2018prediction} found patterns referring to specific days or times (e.g. the time of most accidents, the area with the most consistent number of accidents). However, no attempt was made to understand whether the results are conveyable to non-expert users. 

While road and traffic safety is an important and interesting challenge in the research domain, hence several studies have been focused on pedestrian safety \cite{gaarder2004impact}, \cite{mccomas2002effectiveness}, \cite{mohamed2013clustering}. For example, Mohamed et al., \cite{mohamed2013clustering} used a clustering regression approach to analyze the severity of pedestrian-vehicle crashes in large cities. Their use of the clustering approach helped to examine how the segmentation of the accident datasets help to understand complex relationships between injury severity outcomes and the contribution of the geometric, built environment, and socio-demographic factors.

Previous research studies show a gap in finding pedestrian-oriented accident places. Although, Mohamed et al., \cite{mohamed2013clustering} found that most pedestrian accidents occur in large cities, however, where exactly those accidents occurred in a large city is yet to be a priority in a research study. Moreover, there is a clear gap between research findings and the presentation of those to non-expert users. Perhaps, a study should focus on these aspects and help non-expert users to identify accident-prone areas within a city. Visualization techniques and user study could be a solution to this problem. 

This paper will provide a straightforward visualization based presentation of the findings to the non-expert users and show how the data mining features can significantly help to discover accident-prone areas. In this study, we are going to identify some of the fatal areas of pedestrian road accidents using unsupervised learning algorithms. Then, we will verify the clusters using a heatmap. 

The visualization of statistical analysis will help us more to identify the most dangerous time, day, or month of the year the accidents took place. Also, the road features where the number of accidents is high will give us intuition about the mindset of pedestrians.

\section{Method}
\subsection{Data}
For this study, we downloaded the data from a national database (i.e., \url{https://www.data.gov}) \cite{usdatagov}. The original database contains 71 columns with different attributes. The initial database consisted of 33,707 rows. However, some values were missing. Also, some duplicate values might have occurred due to human errors. Moreover, some of the columns were unnamed. Those columns also had a lot of Not a Number (NaN) values. We used `pandas' (i.e., Python library) data frame to import the data. The database included population-related information such as driver injury, crash locations, the gender of the pedestrian/driver, city/county names, etc.

\subsection{Data Preprocessing}
Initially, we identified important columns that needed to be processed. Then, we considered the most meaningful columns e.g., latitude and longitude. In terms of latitude and longitude, each pair of the columns of each row indicated one crash record. Furthermore, we removed the missing values of  ``latitude'' and ``longitude'', before plotting the coordinates into a map view. Because the main objective was to identify accident-prone places, and missing latitude and longitude values were not helpful to achieve that research goal. Following steps were taken during the data preprocessing:

\begin{itemize}
    \item The ``duplicated().values.any()'' and ``isna().values.any()'' methods of Python programming language helped us to determine if we had any duplicates and missing values in our selected feature columns. 
    \item We again used a method ``dropna()''  that helped us drop the missing values of our ``latitude'' and ``longitude'' columns. 
    \item The duplicate values were not removed because multiple accidents may have occurred in the same location.
\end{itemize}

\subsection{Data analysis}
\subsubsection{Aggregate accidents}
After the preprocessing, we ran a descriptive statistical analysis, and aggregated accidents by days, months, and time to find the number of crashes on each day of each month (Table \ref{tbl:eachday}).

\begin{table}[]
\label{tbl:eachday}
\caption{Describes the number of accidents each day}
\begin{tabular}{lllllllll}
\hline
Months    & Sunday & Monday & Tuesday & Wednesday & Thursday & Friday & Saturday & total cases \\ \hline
January   & 246    & 352    & 469     & 379       & 406      & 438    & 317      & 2607        \\ 
February  & 259    & 327    & 351     & 371       & 374      & 402    & 362      & 2446        \\ 
March     & 268    & 398    & 380     & 374       & 419      & 413    & 402      & 2654        \\ 
April     & 257    & 367    & 386     & 401       & 391      & 425    & 405      & 2632        \\ 
May       & 309    & 333    & 382     & 381       & 445      & 437    & 381      & 2668        \\ 
June      & 287    & 352    & 345     & 328       & 368      & 416    & 373      & 2469        \\ 
July      & 303    & 340    & 363     & 355       & 341      & 383    & 356      & 2441        \\ 
August    & 342    & 378    & 346     & 382       & 382      & 495    & 402      & 2727        \\ 
September & 306    & 408    & 446     & 425       & 414      & 519    & 487      & 3005        \\ 
October   & 340    & 507    & 534     & 578       & 506      & 599    & 511      & 3575        \\ 
November  & 333    & 528    & 489     & 532       & 480      & 508    & 448      & 3318        \\ 
December  & 276    & 452    & 445     & 478       & 503      & 566    & 444      & 3164        \\ \hline
\end{tabular}
\end{table}

\subsubsection{DBSCAN algorithm.}
We used DBSCAN to cluster the co-ordinate points on a leaflet map generated from Python’s ``Folium'' library. With a given set of data points, the DBSCAN clustering algorithm groups together the neighboring points within a given distance \cite{ester1996density}. 

The \textit{DBSCAN} is a density-based clustering algorithm \cite{ester1996density}. It allows us to find $k$ number of incident locations that are with the proximity of a given distance of $\epsilon$. It also allows us to find scattered locations or outliers.  
Suppose there are $n$ number of points (i.e., also referred as a location in this paper), such as $p_1, p_2, \cdot, p_n$. If there are at least $k$ (i.e., the minimum number of points) number of points within a given distance of $\epsilon$, then those points make a cluster together. Also, note that a point $p$ is known to be a core point if and only if there are $k$ points are within $\epsilon$ distance. Points that are not within $\epsilon$ distance and do not form a cluster with $k$  number of points, then those points are called outliers.

To find if a point $q$ is in the neighborhood of point $p$, a distance function $d(p,q)$ can be used, if the distance function returns a value less than or equal to the given distance $\epsilon$ for the point $q$, then $p$ and $q$ form a neighborhood which can be expressed as,

\begin{equation}
    N_\epsilon (p):\{q|d(p,q)\leq \epsilon \}
\end{equation}

In this study, we used the DBSCAN clustering algorithm to identify high accident-prone areas. In this study, $\epsilon$ was 0.05 (50 meters) and $k$ value was 300.

\begin{figure}[h!]
 \centering
 \includegraphics[width=1.0\textwidth]{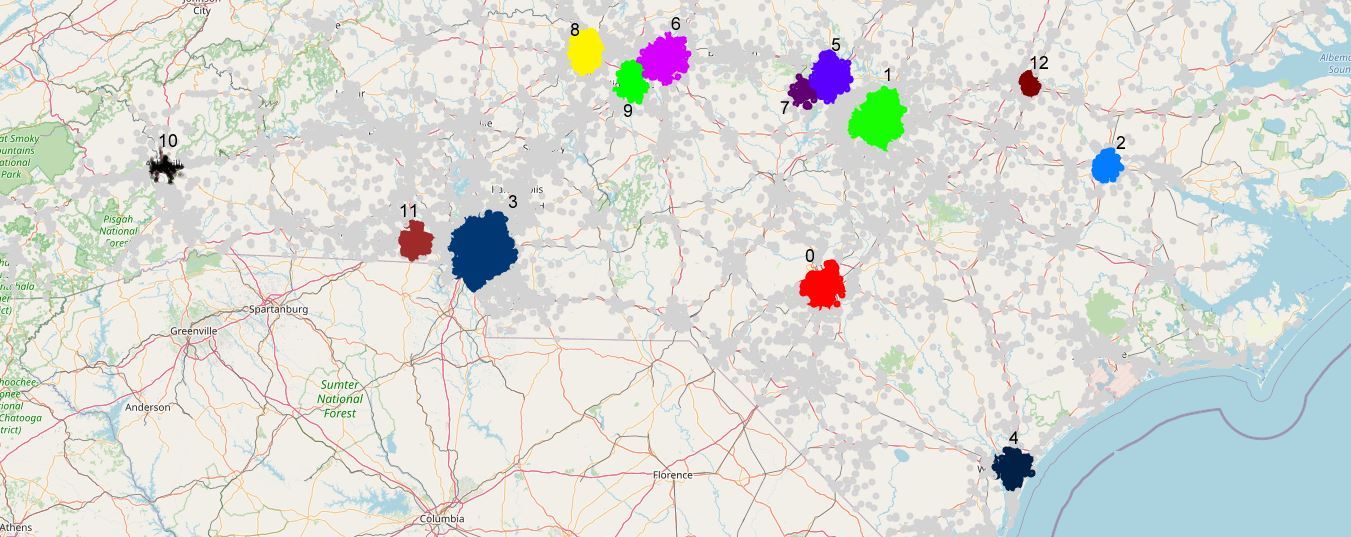}
 \caption{DBSCAN results on the map to show accident-prone areas. There are 13 clusters where at least 300 accidents occurred.}
 \label{fig:dbscan}
\end{figure}

\subsubsection{Heatmap analysis.}
We wanted to make sure that our findings from the DBSCAN clustering algorithm were accurate, so, we tried to find the optimal number of clusters using the ``Silhouette Score''. The \textit{Silhouette Score} helps to explore how many clusters are optimal for a dataset \cite{runfola2013body}. 

There are more methods of measuring the quality of a cluster. For example, Davies- Bouldin Index (DB), the Calinski-Harabasz Index (CH) and Dunn Index etc \cite{bandyopadhyay2002genetic,xu2012comparison}. However, the notion of a ``Good Cluster'' is relative. So, we have further tried to verify the quality of the clusters found in our analysis using a heatmap. The generated heatmap on the map provided us the number of accidents.

To generate the heatmap, we used the Gaussian kernel. The Gaussian kernel provides a better result in heatmap analysis \cite{yu2018face}, which can be expressed as,

\begin{equation}
    e^{-{\frac{(x-x_0)^2+(y-y_0)^2}{\alpha}}}
\end{equation}
where $\alpha$ represents influences of each data points on its surroundings, $x_0$ is the longitude and $y_0$ is the latitude of the current location, and $x$ is the longitude and $y$ is the latitude of a neighboring data point.

\begin{figure}[h!]
 \centering
 \includegraphics[width=1.0\textwidth]{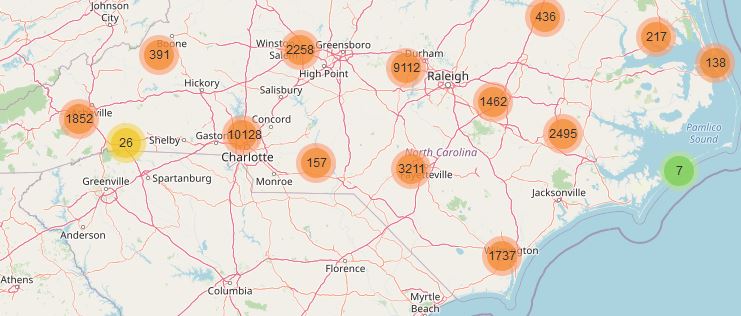}
 \caption{The ``Marker Cluster'' is showing us the number of accidents in our region of interest. We could see clusters nearly in similar areas of Fig. \ref{fig:dbscan}.}
 \label{fig:markercluster}
\end{figure}

The generated heatmap complements our findings from the clusters (see Fig. \ref{fig:dbscan} and \ref{fig:heatmap}). It also verifies that the quality measurement (Silhouette Score) of the clusters showing that the quality was good. The map gives us an idea about some of the densest areas of the accidents, as well as the numbers, represent the number of crashes happening around that area. Also, as we compared our results, and found that the numbers of the accidents correspond to each of the clusters using marker cluster as well (See Fig. \ref{fig:markercluster} and \ref{fig:heatmap}). 

The marker cluster were generated using following equation,
\begin{equation}
    (\sqrt{(x_0\times C-x\times C)^2 + (y_0\times C-y\times C)^2})>>(Z_{m}-Z_{c})
\end{equation}
where $C$ is a constant, $x_0$ is the longitude and $y_0$ is the latitude of the current location, and $x$ is the longitude and $y$ is the latitude of a neighboring data point. $Z_m$ is the maximum zoom value (predefined to 22), and $Z_c$ is the current zoom value.


\begin{figure}[h!]
 \centering
 \includegraphics[width=1.0\textwidth]{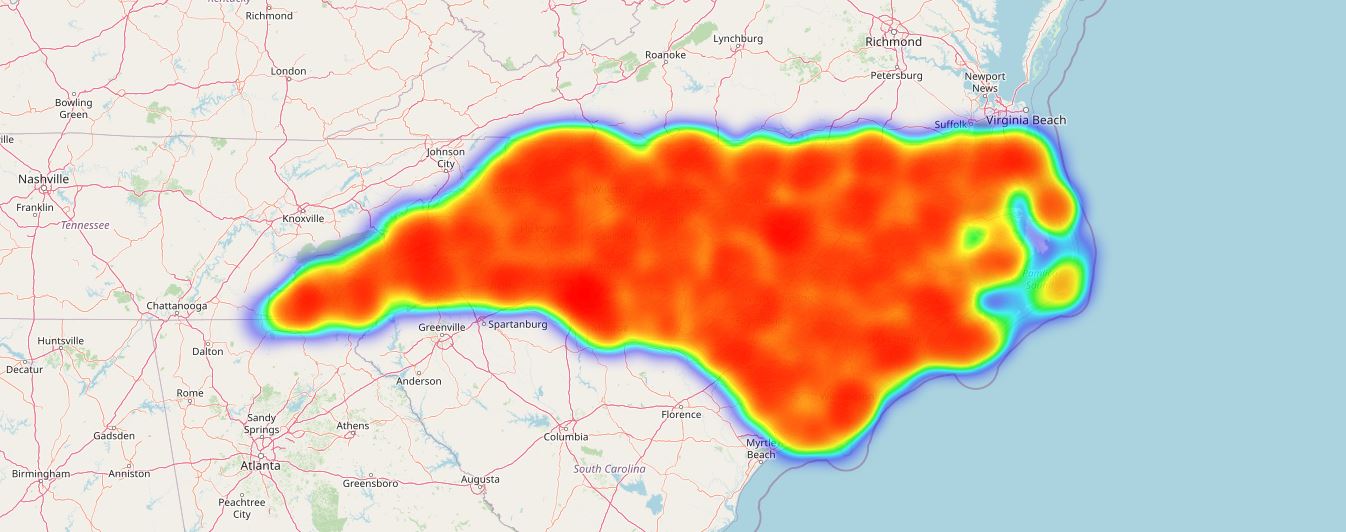}
 \caption{Heatmap to show concentration of incidents in an area}
 \label{fig:heatmapdistance}
\end{figure}

\begin{figure}[h!]
 \centering
 \includegraphics[width=1.0\textwidth]{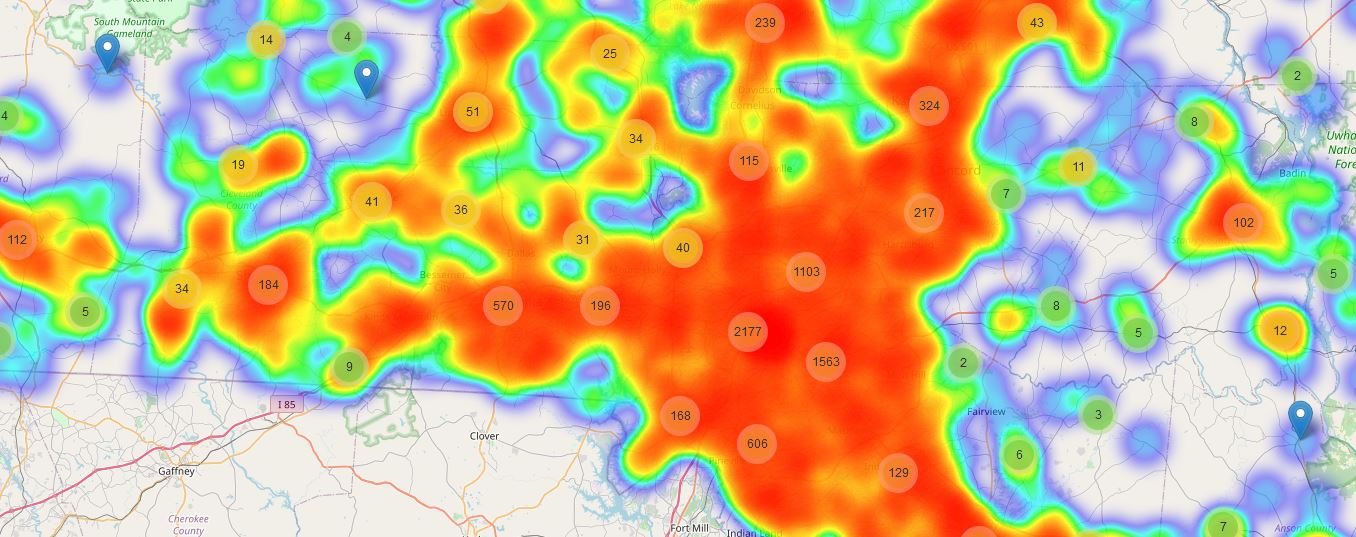}
 \caption{Heatmap to show concentration of incidents in an area. This figure is a zoomed version of Fig. \ref{fig:heatmapdistance}}
 \label{fig:heatmap}
\end{figure}

\subsubsection{Analysis of visualization techniques.}
Finally, we tried to extract some key points from the research data that can provide us a broad picture of the accidents (e.g. The months/days/time of the highest number of crashes or the road features of the crashes). We tried to represent the results by visualizing it with advanced, yet very simple to interpret visualization techniques. We took a survey of 50 participants who are not involved in the visualization development process and tried to find out the best visualization techniques among several graph/pie charts representations. We also tried to see if non-expert users could interpret the data from the figures (see Fig. \ref{fig:accidenttime}, \ref{fig:accidentday}, and \ref{fig:accidentroad}).

\section{Results and discussions}
On the marker cluster map, the cluster with 10,128 accidents indicated a significantly higher number of accidents than other clusters (see Fig. \ref{fig:markercluster}). Similarly, our results indicated that another cluster with 9,112 number of accidents. The marker cluster map presents all the accidents grouped with their locations and neighboring location as well. Figure \ref{fig:markercluster} indicates that most of the accidents took place almost in the middle of the city. Perhaps this was the case because the population density in the middle of the city is comparatively higher than the rest of the places. 

The Heatmap we generated was a two-layer visual map. The first layer was the ``Heat'' and the second one was the marker. If we enable both layers then we get a better picture of the accidents-prone areas (See Fig. \ref{fig:heatmap}). When we zoom in the map or select a specific cluster, then cluster areas with detailed information will appear. The numbers are broken down into smaller ones. Also, the heatmap view could help to understand which areas are more accident-prone than others. These visualization results could help non-expert users to decide in which areas they have to be more cautious or pay more attention.

\subsection{Crash analysis by month} 
\begin{figure}[h!]
 \centering
 \includegraphics[width=1.0\textwidth]{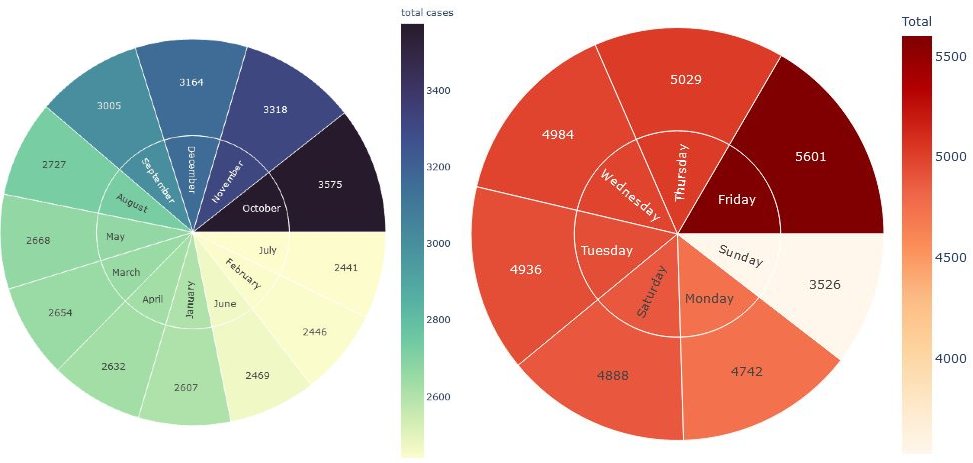}
 \caption{(Left) The Sunburst chart gives us a depiction of the month with the most number of accidents, October. (Right) Showing us the day with the most number of accidents. Friday has the most number of total cases with over 5,600 crashes.
}
\label{fig:sunburst}
\end{figure}

The most number of crashes have happened during the month, October (Fig. \ref{fig:sunburst} (left)). Our results indicated that more than 3,500 accidents occurred during this period. One of the plausible explanations for this case is that October lies in the middle of autumn in North Carolina \cite{boyles2003analysis}. The second most number of accidents have taken place in the month, November with 3,318 cases. Then comes the month December with 3,164 filed cases of accidents. The month where the least number of accidents have happened is July with 2,441 cases. July is usually the warmest weather of the year in North Carolina and is the rainiest month in Chapel Hill. July month is in the middle of the summer season in North Carolina \cite{boyles2003analysis}. So, perhaps the presence of natural light helps to prevent accidents in July.

\subsection{Crash analysis by time} 
\begin{figure}[h!]
 \centering
 \includegraphics[width=0.7\textwidth]{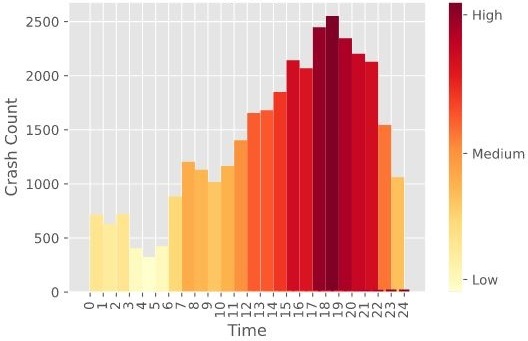}
 \caption{Showing us accidents each hour. The most number of accidents have happened between hours 18 (i.e., 6 pm) and 19 (i.e., 7 pm).}
 \label{fig:accidenttime}
\end{figure}

The number of accidents happening on each hour could be explained by the Figure \ref{fig:accidenttime}. The most number of accidents have taken place between hours 18 and 19. More than 2,500 accidents happened during that time. The least number of accidents have taken place during the midnight and post-midnight hours, from the 0th hour to 6th. The higher number of accidents occurred starting from the 15th hour till the evening at 21st. The least number of accidents happened within the hour 4 and 5 (see Fig. \ref{fig:accidenttime}). This could be because this is the time most people leave their workplace and rush to go home. As a result, most accidents occur during the hours of 18 and 19. Similar findings were reported by Amiruzzaman \cite{amiruzzaman2018prediction}, where the author found most traffic accidents occurred in DC after office hours.

\subsection{Crash analysis by day}
\begin{figure}[h!]
 \centering
 \includegraphics[width=1.0\textwidth]{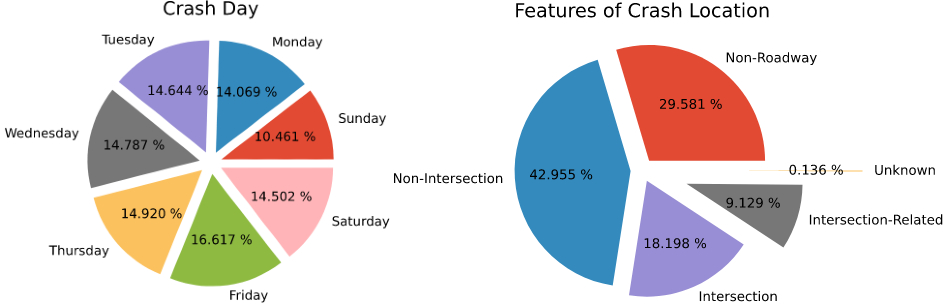}
 \caption{(Left) The exploded pie chart is showing us the most number of accidents on each day of the week by percentage. (Right) The exploded pie chart is showing us the most number of accidents happening on the ``Non-intersection'' roads with 42.955\% of the total accidents.}
 \label{fig:accidentday}
\end{figure}

The percentage of accidents happening on different days of the week are shown in Fig. \ref{fig:sunburst} (right) and Fig. \ref{fig:accidentday} (left).  The evidence from the results which is presented in the exploded pie chart (Fig. \ref{fig:accidentday} (left)) and the sunburst (Fig. \ref{fig:sunburst} (right)) show that most of the accidents have taken place on Fridays. This may be because Friday is the end of weekdays in North Carolina, and most people are eager to relax and pays less attention. Evidence suggested that more than 5,600 crashes or 16.617\% of the total crashes have happened during the Fridays. It is surprising that the most number of accidents happening on the last day of the working week. The day with the least number of accidents is Sunday with 10.461\% of the total accidents. Similar findings were reported in a previous study \cite{smith2004traffic}, our findings confirm those findings.

\subsection{Crash location features} 
\begin{figure}[h!]
 \centering
 \includegraphics[width=0.7\textwidth]{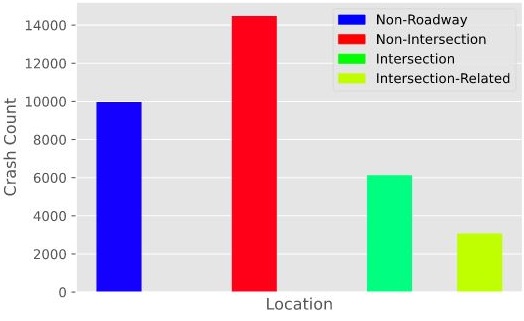}
 \caption{Showing us the number of accidents happening on different roads. Non-intersections have more than 14,000 accidents. The intersection, in comparison, has the least number of accidents with a number of around 3,000.}
 \label{fig:accidentroad}
\end{figure}

It is an astounding fact that the highest number of accidents has emerged from the ``Non-Intersections'' (Fig. \ref{fig:accidentday}(right), and \ref{fig:accidentroad}). The number is seemingly higher than the other ones. More than 14,000 accidents took place at the non-intersection locations. The second place where the accident numbers are high is where the ``Non-Roadways'' are. Around 10 thousand accidents took place. The ``Intersection'' location is where about 6 thousand accidents took place. It is a bit surprising fact too because usually, the intersections points are where the number of transports could be found even more.







\section{Conclusion}
Obtained results from data mining and data analysis suggested that there is a higher rate of accidents in the middle of the city. The most number of accidents have occurred in Charlotte. Also, the most dangerous time to go out is within the hour 18 and 19. The accident numbers are significantly higher than usual during the time between the afternoon and evening, from 3:00 pm to 8:00 pm. We also found that the most number of accidents happened in October, in the middle of autumn. And a higher percentage of accidents happened on Fridays. 

The most number of accidents happened on the Non-intersection roads and the least number of accidents happened on Intersections. It can be said that people tend to be less cautious about the environment if it is not an intersection and increase the chances of getting involved in crashes. However, this tendency needs to be changed. Overall, people should be more cautious if they are going out during these specific times.

Perhaps, analyzing more data from the updated database from the law-enforcement agencies could help us to find more interesting information. Also, more data mining techniques could be used to predict the accident-prone regions. As for the future study, we can suggest that different supervised learning algorithms can be used to predict the accident-prone areas. Moreover, we can use deep learning techniques to classify areas based on other features as well.

\section*{Acknowledgement}
We would like to thank the Institute of Energy, Environment, Research, and Development (IEERD, UAP) and the University of Asia Pacific for financial support.

\bibliographystyle{splncs04}
\bibliography{mybibliography}

\end{document}